\documentclass[prb,citeautoscript,twocolumn,groupedaddress]{revtex4-1} 
\usepackage{graphicx}
\usepackage{units}
\usepackage[intlimits]{amsmath}
\usepackage{amssymb}
\usepackage{hyperref}
\usepackage{siunitx}
\usepackage{multirow} 
\usepackage{siunitx} 
\usepackage{natbib}
\usepackage{comment}

\usepackage{hyperref}
\usepackage{amsmath}

\begin{document}

\title{Observation of squeezed light in the 2\,$\mathrm{\mu m}$ region}

\author{Georgia~L.~Mansell$^{1,2,3*}$, Terry~G.~McRae$^{1}$, Paul~A.~Altin$^{1}$, Min~Jet~Yap$^{1}$, Robert~L.~Ward$^{1}$, Bram~J.J.~Slagmolen$^{1}$, Daniel~A.~Shaddock$^{1}$,  and David~E.~McClelland$^{1}$}

\address{
$^{1}$OzGrav, Department of Quantum Science, Research School of Physics and Engineering, Australian National University, Acton, Australian Capital Territory 2601, Australia\\
$^{2}$LIGO Hanford Observatory, P.O. Box 159, Richland, Washington 99352, USA\\
$^{3}$Massachusetts Institute of Technology, Cambridge, Massachusetts 02139, USA\\
$^*$Corresponding author: georgia.mansell@anu.edu.au
}

\begin{abstract}
We present the generation and detection of squeezed light in the 2\,$\mathrm{\mu m}$ wavelength region. This experiment is a crucial step in realising the quantum noise reduction techniques that will be required for future generations of gravitational-wave detectors. Squeezed vacuum is generated via degenerate optical parametric oscillation from a periodically-poled potassium titanyl phosphate crystal, in a dual resonant cavity. The experiment uses a frequency stabilised 1984 nm thulium fibre laser, and squeezing is detected using balanced homodyne detection with extended InGaAs photodiodes. We have measured $4.0 \pm 0.1$ dB of squeezing and  $10.5 \pm 0.5$ dB of anti-squeezing relative to the shot noise level in the audio frequency band, limited by photodiode quantum efficiency. The inferred squeezing level directly after the optical parametric oscillator, after accounting for known losses and phase noise, is 10.7 dB.
\end{abstract}

\maketitle



\section*{Introduction}
	The era of gravitational-wave astronomy began with the detections of four binary black hole mergers by the Advanced Laser Interferometer Gravitational-Wave Observatory (aLIGO) detectors\cite{LSC150914, LSC151226, LSC170104, LSC170608},  a co-incident binary black hole detection with the advanced Virgo detector \cite{LSC170814}, and, in August 2017, a binary neutron star coalescence observed by aLIGO, Virgo, and subsequently by many electromagnetic telescopes.  This last event launched a new type of multi-messenger astronomy \cite{LSC170817, LSC170817MM}. 

Despite these recent breakthroughs, gravitational-wave astronomy will be limited by the sensitivity of the detectors for the foreseeable future.  Even when aLIGO reaches design sensitivity, the detection horizon will remain limited to the nearby universe. Fully exploiting the potential of gravitational-wave astronomy will require detectors with cosmological reach.  In particular, black hole spectroscopy and making inferences about the population of binary black holes from Population III stars will require at least ``LIGO-Voyager" class detectors \cite{Berti2016,Belcynski2017}.  LIGO-Voyager \cite{ISWhitepaper2015} is a proposed new detector in the existing LIGO facility, designed to fully exploit the limits of the facility (set primarily by the large ultra-high-vacuum system). LIGO-Voyager has a target binary neutron star range of 1100 Mpc and will detect GW150914-like binary black holes to a redshift of $z \sim 7$, vs $z \leq 2$ for aLIGO \cite{VoyagerDesign}; binary black holes from Population III stars may contribute a significant fraction of the merger rate density from about $z \gtrsim 8$ \cite{Belcynski2017}. 

LIGO-Voyager, and other future interferometric gravitational wave detectors plan to use new test mass materials, with reduced loss and scatter to improve on thermal and scatter noise, such as crystalline silicon. The proposed LIGO-Voyager upgrade aims to optimise the detector sensitivity, with upgrades to test masses, suspensions, optical components and laser systems, while using the current LIGO sites and vacuum envelope. The current LIGO-Voyager design utilises 100 kg silicon test masses cooled to 123 K, to exploit the low mechanical loss and zero thermal expansion coefficient of the bulk material. The baseline wavelength of the design is in the 2\,$\mathrm{\mu m}$  region, to make use of the low coating absorption (relative to 1550\,nm \cite{CoatersDozen}), and reduced optical loss and scatter properties of silicon, the latter of which reduces with $1/\lambda^2$, where $\lambda$ is the operating wavelength of the laser interferometer. 

Quantum noise in GW detectors is driven by vacuum fluctuations, which couple to the interferometer via the dark readout port.  Injection of audio-band squeezed vacuum states into the interferometer dark port is a well-known quantum noise reduction technique \cite{Caves1981,Braginsky2001c}. Squeezing injection has been demonstrated in prototypes \cite{Goda2008} and on the previous generation of gravitational wave detectors at both the GEO-600 \cite{LIGONPMay2011} and LIGO Hanford \cite{LIGONPJuly2013} observatories, and is currently being installed on the advanced gravitational wave detectors. 
		
To achieve design sensitivity, LIGO-Voyager requires broadband quantum noise reduction through the injection of 10\,dB squeezed light.  The results presented here are the first step towards a 2\,$\mathrm{\mu m}$ squeezer for LIGO Voyager, and an important pathfinder technology for this and other future detectors. 

While this research is motivated by future gravitational wave detectors, 2\,$\mathrm{\mu m}$ lasers are also 	used in other applications such as LIDAR, gas sensing, free-space optical communications, medicine, and materials processing \cite{2umLaser, HoDopedFibreLaser, HoDopedSilicaReview}, which can all benefit from the kind of high sensitivity, low noise measurements presented here.

\section*{The experiment} 

\begin{figure}
	\centerline{\includegraphics[width=80 mm]{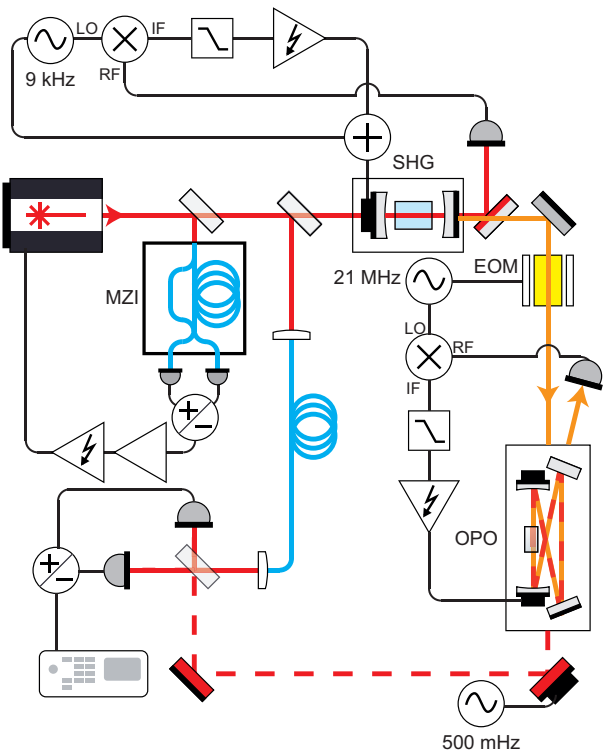}}
	\caption{Schematic of the squeezer, 1984 nm beams shown in red, 992 nm beams in orange, electrical connections in black, and optical fibre in blue. The laser is stabilised to a passive fibre Mach-Zehnder interferometer (MZI). The laser is frequency doubled at the second harmonic generator (SHG) to produce the 992 nm pump beam for the optical parametric oscillator (OPO). The OPO is locked to the pump beam via Pound-Drever-Hall (PDH) locking on reflection \cite{PDH1983}. To generate squeezed vacuum the OPO is locked at a dual resonance point at the crystal phase matching temperature. The squeezed beam is reflected off multiple highly reflective dichroic beamsplitters, to remove residual pump light, and mixed with a fibre-delivered local oscillator. The shot noise, squeezing and anti squeezing are then detected using balanced homodyne detection.}
	\label{fig:2umExpt}
\end{figure}
	There are significant challenges involved with operating an experiment at 2\,$\mathrm{\mu m}$, as technologies in this wavelength region are less developed. Lasers at this wavelength have poorer frequency and intensity noise, and poorer overall stability, compared to their lasers used for squeezed light generation at 1064 nm. Optics, detectors, and modulators are more difficult to obtain and are more expensive, and detectors with high quantum efficiency and low dark noise are currently under development. For this demonstration of squeezed light at 2\,$\mathrm{\mu m}$, we have used basic laser frequency stabilisation techniques.
	
	Precise choice of laser wavelength within the ``2\,$\mathrm{\mu m}$ region" was based on a window of high transparency in the atmospheric absorption spectrum \cite{ARTRAN, gemini}, which also corresponds to peak emission of the thulium lasing ion \cite{Zou1996}. Other design choices, including the optical and mechanical layouts of the two nonlinear optical cavities in the experiment, followed the basic design of previous squeezed light experiments  \cite{Chua2011, Wade2015}.

	Figure \ref{fig:2umExpt} shows the experimental layout, which utilises a 2 $\mathrm{W}$ thulium fibre laser and amplifier from AdValue Photonics. The laser has a 50 kHz linewidth, and is frequency stabilised to a passive thermally-isolated fibre Mach-Zehnder interferometer (MZI)  with a 3 m arm length mismatch, to reduce slow laser frequency drift. To lock the laser to the MZI, a simple error signal is generated using balanced homodyne detection on both outputs of the MZI. The error signal is filtered and fed back to the laser piezoelectric transducer (PZT) for frequency actuation. 
	
	A tap-off of the main laser is frequency doubled to 992~nm in an external second harmonic generator (SHG) to produce the pump light for the squeezer optical parametric oscillator (OPO). The SHG consists of a $1 \times 1 \times 10$~mm periodically poled potassium titanyl phosphate (PPKTP) crystal, in a linear singly-resonant cavity. Both faces of the crystal which are encountered by the beam are coated with anti-reflective dielectric coatings. The cavity consists of two mirrors both with -15 mm radius of curvature, with one attached to a PZT for cavity length actuation. The input coupler has a reflectivity of 94\% at 1984 nm and $>$99.9\% at 992 nm, while the output coupler has a reflectivity of $>$99.9\% at 1984 nm and $<$5\% at 992 nm. The cavity has a finesse of approximately 100 (at the pump wavelength), a linewidth of 36 MHz, and a physical length of 33 mm (the optical path length is wavelength dependent due to the index of refraction in the crystal). To lock the SHG on resonance with the laser, the cavity length is dithered at 9 kHz which is demodulated on transmission, filtered and fed back to the cavity length. The dither lock provides an inexpensive alternative to Pound-Drever-Hall (PDH) locking, with no phase modulator required; a bandwidth of 125 Hz was achieved with this technique. Transmitted light at the fundamental wavelength is removed with dichroic mirrors, and the 992 nm pump beam is mode matched to the OPO. The SHG can produce up to 200 mW of pump power.
	
	The OPO is a dual resonant bow-tie design following previous ANU squeezers \cite{Chua2011, Wade2015,LIGONPJuly2013}.  The PPKTP nonlinear crystal ($1 \times 5 \times 12 $ mm) has a $4 ^\circ$ wedge at one end. To achieve dual resonance while the crystal is at the optimal phase matching temperature, we adjust the lateral position of the crystal until the differing optical path lengths through the crystal for the pump and fundamental compensate for the cavity dispersion.  The angle of the crystal wedge is chosen such that there are 5 dual resonance paths, with respect to the 1984 nm beam, through the crystal, so that a path free of crystal defects can be selected. The cavity parameters are shown in table \ref{table:2umOPO}. Parameters were chosen for optimal escape efficiency, threshold, cavity stability, beam waist size, and higher order mode spacing. The OPO is locked to the laser using the Pound-Drever-Hall technique on reflection of the pump beam using phase modulation sidebands at 21 MHz. 
	
\begin{center}
	\begin{table}
		\begin{tabular}{ l | l }
		Parameter & Value \\ \hline
		Cavity length & 459 mm\\
		Mirror radius of curvature & -50 mm\\
		$\omega_0$ in crystal (1984 nm)& 28.5 (s) 30.2 (t) $\mu $m\\
		$\omega_0$ external (1984 nm)& 512.59 (s) 463.05 (t) $\mu $m\\
		Finesse (1984 nm) & 38.6 \\
		Linewidth (1984 nm) & 16.54 MHz \\
		Free spectral range (1984 nm) & 639 MHz \\
		Input coupler reflectivity (1984 nm) & 85\% \\
        Crystal poling length & 36.525 $\mu $m \\
        \hline
		Escape efficiency, $\eta_{esc}$ (1984 nm) & $96.7 \pm 0.1 \%$ \\
		Threshold power (992 nm) & 315 mW \\
        Phase matching temperature & 33.8 C
		\end{tabular}
		\caption{Table of measured (lower section) and modelled (upper section) parameters for the squeezer optical parametric oscillator.}
		\label{table:2umOPO}
	\end{table}
\end{center}

	PPKTP has been used to demonstrate squeezing at 1064 nm and 1550 nm \cite{Mehmet2011} in past experiments, and has a transparency range between 400 - 4000 nm \cite{Valentin1999}. The nonlinear crystal has a broader phase matching temperature acceptance compared to its 1064 nm counterpart due to a lower refractive index difference between the pump and fundamental at this longer wavelength. The nonlinearity for this 1984 nm experiment was measured directly in a single pass SHG measurement to be $d_{eff} = 6.4$~pm/V, which is lower than the expected value, assuming the nonlinearity is consistent between 1064 nm and 1984 nm. The nonlinearity of PPKTP has been measured to be 9.3 pm/V at 1064 nm \cite{Shoji1997}, and reported as 7.3 pm/V at 1550 nm \cite{Ast2013}. The measured threshold of the OPO cavity is also significantly higher than expected given modelled cavity parameters, which is consistent with reduced nonlinearity. A precise characterisation of the nonlinearity and loss of the crystal is ongoing.
	
	Squeezed vacuum is produced in the OPO, and propagates to the balanced homodyne detector. At the homodyne detector  the squeezed beam is combined with a local oscillator tapped off from the main laser and delivered via optical fibre, which also acts as a spatial mode cleaner. The homodyne detector uses two 500\,$\mathrm{\mu m}$ diameter extended InGaAs photodiodes with windows removed and measured quantum efficiencies of $\eta_{QE} \sim 74 \pm 5\%$. Other detector materials were considered, including mercury cadmium telluride, however extended InGaAs was chosen for its relatively high detectivity, low dark noise, and room-temperature operation. Through careful optimisation of the beamsplitter angle, and beam position on the photodiode, the common mode rejection of the local oscillator path on the homodyne detectors was measured to be 70 dB. The diodes are operated with no bias voltage to reduce the dark current, which otherwise would be a limiting noise source. This mode of operation limits the detector bandwidth to approximately 500 kHz.
		 	
\section*{Results}

	\begin{figure}
	\centerline{\includegraphics[width=80 mm]{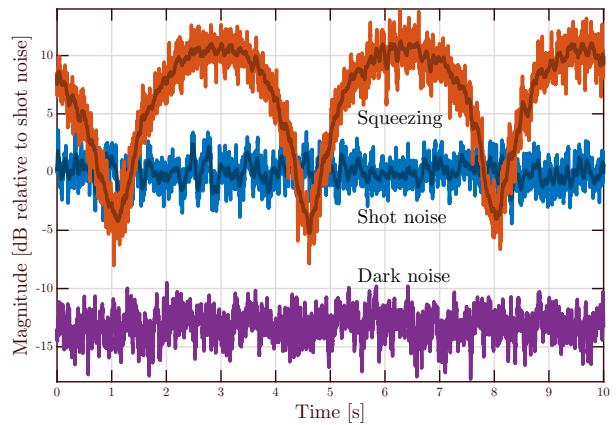}}
	\caption{Squeezing level as a function of time, while scanning the local oscillator phase, normalised to the local oscillator shot noise. This is a zero-span measurement at 30 kHz with a resolution bandwidth of 1 kHz and a video bandwidth of 30 Hz. The shot noise is shown in blue, squeezing arches in red, and homodyne detector dark noise in purple. A moving average over every 50 data points is shown in darker colors; that the total data set has 1715 points. Data taken using an Agilent E4402B spectrum analyser.}
	\label{fig:sqzArches}
\end{figure}

\begin{figure}
	\centerline{\includegraphics[width=80 mm]{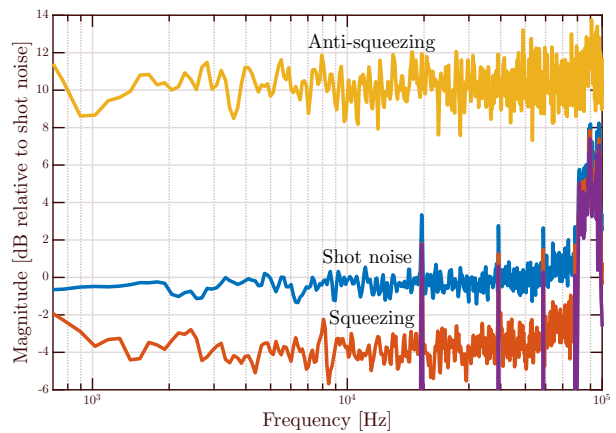}}
	\caption{Spectra of squeezing, shot noise and anti-squeezing, normalised to the shot noise level, taken under the same conditions as figure \ref{fig:sqzArches}. High frequency roll up in all spectra is due to the homodyne detector dark noise, shown in purple. Low frequency roll-up on the squeezing and anti-squeezing paths is due to stray light on the squeezing path.}
	\label{fig:sqzSpectrum}
\end{figure}

	The results from the squeezer are shown in figures \ref{fig:sqzArches} and \ref{fig:sqzSpectrum}; we have measured $4.0 \pm 0.1$\,dB of squeezing and $10.5 \pm 0.5$\,dB of anti-squeezing in the audio band. Figures \ref{fig:sqzArches} and  \ref{fig:sqzSpectrum} show respectively a zero span measurement at 30 kHz, and a spectrum from 700 Hz to 100 kHz. The local oscillator power was roughly 100 $\mathrm{\mu W}$ on each photodiode, this allowed for 13 dB of dark noise clearance, measured with the optical path from the squeezer OPO blocked. To measure the magnitude of squeezing and anti-squeezing the phase of the local oscillator was scanned at 0.5 Hz using a PZT mounted mirror, which dithers the measurement quadrature. 
	
	The measured squeezing and anti-squeezing levels, expressed as variances, are fully explained by the loss, noise in the squeezing ellipse phase, and the nonlinear gain produced by the crystal \cite{Dwyer2013}. The variances in each quadrature, in the absence of phase noise, can be written as

	\begin{equation}
		V_{1,2} = \eta V^{in}_{1,2} + 1 - \eta
	\end{equation}
	
	 where $V_1$ is the variance in the anti-squeezed quadrature, $V_2$ is the variance in the squeezed quadrature, $\eta$ is the measurement efficiency, and $V^{in}_{1,2}$ are the variances in each quadrature, before the sources of loss. Phase noise, which is jitter on the squeezing ellipse, introduces coupling between the squeezed and anti-squeezed quadratures. Assuming the phase noise, $\delta \theta$, is small, the variances become:
	 
	  \begin{equation}
		V_{1,2}(\delta \theta) = \cos^2 \delta \theta \ V_{1,2} + \sin^2 \delta \theta \ V_{2,1}.
	\end{equation}
	 
	 The measurement efficiency is calculated from the collective losses of the system, including contributions from the OPO cavity escape efficiency, the fringe visibility, propagation from the OPO to the homodyne, and the photodiode quantum efficiency. The propagation loss on the path from the OPO to the homodyne was measured to be \textless 1\%. The homodyne fringe visibility, a measure of the overlap between the squeezed beam and the local oscillator, was measured to be $\nu_{vis} = 98\pm1.3\%$, and $\eta_{vis} = (\nu_{vis})^2$. The escape efficiency, a measure of the out-coupling efficiency of the OPO, was $96.7 \pm 0.1 \%$, and the photodiode quantum efficiency was $\eta_{QE} = 74 \pm 5\%$. Thus the total efficiency is 
	 
	 \begin{equation}
	 \eta = \eta_{esc} \times \eta_{prop} \times \eta_{vis} \times \eta_{QE} = 69 \pm 7 \%.
	 \end{equation}
	 
	A precise measurement of the squeezing ellipse phase noise was precluded by the low nonlinear gains achievable with the crystal. The phase noise was estimated by measuring the phase fluctuations between the local oscillator and a diagnostic bright beam at the fundamental wavelength that was transmitted through the OPO cavity. This yielded a value of 50\,mrad, predominantly caused by a 300\,kHz oscillation in the laser. The injected pump power was set to give a nonlinear gain of $7 \pm 0.5$, higher gains would result in lower measurable squeezing levels due to the squeezing ellipse phase noise. To achieve 10\,dB of squeezing with 50\,mrad of phase noise places unrealistic requirements on the loss of the measurement, as shown in Dwyer et al. \cite{Dwyer2013} and Eqs. (1) and (2), with the total loss required to be \textless 0.2\%. We see no technical reason why phase noise as low as 1.3\,mrad, as demonstrated at 1064 nm \cite{Oelker2016} cannot be reproduced at 2\,$\mathrm{\mu m}$  in a future experiment, thus increasing the permissible loss to measure 10 dB of squeezing up to 10\%.

	Given the loss, phase noise, and nonlinear gain, we calculate the expected squeezing level to be $4.2 \pm 0.7$~dB and antisqueezing level to be $11.1\pm0.3$~dB. We infer the level of generated squeezing to be 10.7 dB by removing sources of loss and phase noise from this model.
	
	The reduced crystal nonlinearity increases the threshold of the cavity, thus higher pump power levels are required to reach a nominal nonlinear gain. While this requires higher power lasers and deposits additional heat in the nonlinear crystal, it does not limit squeezing levels in our experiment, or prevent access to the level of squeezing required by LIGO Voyager. To increase circulating power, the finesse of the OPO cavity can also be increased.
		
	

\section*{Conclusions and future work}
	We have made the first observation of squeezed light in the 2\,$\mathrm{\mu m}$ region. This experiment paves the way for future measurements beyond shot noise at 2\,$\mathrm{\mu m}$. We have demonstrated that 2\,$\mathrm{\mu m}$ is a viable option for the operating wavelength of future gravitational wave detectors, and that 1984 nm is a promising candidate for the specific wavelength in the 2\,$\mathrm{\mu m}$ region. While this experiment is not limited by the crystal nonlinearity, further investigation into nonlinear crystal behaviour is ongoing, seeking to determine whether the measured low nonlinearity is peculiar to the crystals used in this work or due to the properties of PPKTP at 2\,$\mathrm{\mu m}$. This will help to optimise future squeezer OPO optical design. 
	
	To achieve higher levels of squeezing the issues of phase noise and loss must be addressed. The high phase noise in this experiment can be mitigated with improvements to laser stability, which we believe is achievable with current technology. To improve on the loss, the quantum efficiency of photodetectors in the 2\,$\mathrm{\mu m}$ region must be improved. This requires further advances in semiconductor materials, however there is no fundamental reason why quantum efficiency photodiodes cannot be made for 2\,$\mathrm{\mu m}$ \cite{Uwe2016}. Future developments for this experiment will also include a control scheme to lock the squeezing angle, which will allow squeezing to be observed over the full ground-based gravitational-wave detector band.
		
\section*{Acknowledgements}
We acknowledge discussions with Rana Adhikari, and thank the MIT LIGO lab for contributing to the homodyne detector. This research was supported by the Australian Research Council under the ARC Centre of Excellence for Gravitational Wave Discovery, grand number CE170100004. B.S. has been supported by ARC Future Fellowship FT130100329.

---
%
%
%


\bibliographystyle{unsrt}

\bibliography{library}


\end{document}